\documentclass{iaus}
\usepackage{graphicx}

\title[Finding the sub-stellar companion candidate CT Cha] 
{Finding new sub-stellar co-moving companion candidates - the case of CT Cha}

\author[Tobias Schmidt, Ralph Neuh{\"a}user
]   
{Tobias Schmidt$^1$
 \and Ralph Neuh{\"a}user$^1$
}

\affiliation{$^1$Astrophysikalisches Institut und Universit\"ats-Sternwarte, Universit\"at Jena, \\ Schillerg\"a\ss chen 2-3, 07745 Jena, Germany \\ email: {\tt tobi@astro.uni-jena.de} \\[\affilskip]
}

\pubyear{2008}
\volume{249}  
\pagerange{}
\jname{Exoplanets: Detection, Formation and Dynamics}
\editors{A.C. Editor, B.D. Editor \& C.E. Editor, eds.}
\begin{document}

\maketitle

\begin{abstract}
We have searched for close and faint companions around T Tauri stars in the Chamaeleon star forming region. Two epochs of direct imaging data were taken with the VLT Adaptive Optics instrument NaCo in February 2006 and March 2007 in Ks band for the classical T Tauri star CT Cha together with a Hipparcos binary for astrometric calibration. Moreover a J band image was taken in March 2007 to get color information.
We found CT Cha to have a very faint companion (Ks$_{0}$\,=\,14.6 mag) of $\sim$\,2.67'' separation corresponding to $\sim$\,440\,AU. We show that CT Cha A and the faint object form a common proper motion pair and that the companion is not a non-moving background object (with 4\,$\sigma$ significance).
\end{abstract}

\firstsection 
\section{Introduction}

CT Cha introduced in the 65th Name-List of Variable stars by \cite[Kholopov et al. (1981)]{Kholopov81}, was originally found by \cite[Henize \& Mendoza (1973)]{Henize73}, often called HM\,9, an emission-line star in Chamaeleon exhibiting variations in its H$\alpha$\,line from plate to plate and partial veiling \cite[(Rydgren 1980)]{Rydgren80}.

While the star was first classified as T Tauri star by \cite[Whittet et al. (1987)]{Whittet87} it was later
on found to be a classical T Tauri star by \cite[Weintraub (1990)]{Weintraub90} and \cite[Gauvin \& Strom (1992)]{Gauvin92} from IRAS data, before \cite[Natta et al. (2000)]{Natta00} could find evidence for a silicate feature
disk with $L_{sil}=10^{-2}\,L_{\odot}$ and $L_{sil}/L_{\ast}=0.014$ using ISO data.

The variations of the H$\alpha$\,line could later be interpreted when \cite[Hartmann et al. (1998)]{Hartmann98} measured a mass accretion rate of log{\,\.{M}}$=-8.28\,M_{\odot}/yr$. Additional variations of infrared \cite[(Glass 1979)]{Glass79} and optical photometry can possibly be explained by surface features on CT Cha and a possible rotation period of 9.86 days found by \cite[Batalha et al. (1998)]{Batalha98}.

All additional properties of the K7 \cite[(Gregorio Hetem et al. 1988)]{Gregorio88} star CT Cha, such as e.g. its age estimated to be 3 Myr by \cite[Feigelson et al. (1993)]{Feigelson93} or 0.9 Myr by \cite[Natta et al. (2000)]{Natta00} as well as its equivalent width of the lithium absorption line of W$_{\lambda}$(Li)\,= 0.40\,$\pm$\,0.05\,\AA\,\cite[(Guenther et al. 2007)]{Guenther07} and its radial velocity of 15.1\,$\pm$\,0.1 km/s \cite[(Joergens 2006)]{Joergens06} and proper motion (table\,\ref{tab1}) are consistent with a very young member of the Cha I star-forming cloud with an age of 2\,$\pm$\,2\,Myr and a medium radial velocity of 14.9\,$\pm$\,1.7 km/s \cite[(Melo et al. 2003)]{Melo03}.

\section{Direct observations of a wide companion}

\subsection{AO imaging detection}

We observed CT Cha in two epochs in February 2006 and in March 2007 with the European Southern Observatory (ESO) Very Large Telescope (VLT) at Cerro Paranal in the programmes with IDs 076.C-0292(A) \& 078.C-0535(A). All observations were done using the Adaptive Optics (AO) instrument Naos-Conica (NaCo, \cite[Lenzen et al. 2003]{Lenzen03} \& \cite[Rousset et al. 2003]{Rousset03}).
In all cases the S13 camera ($\sim$13 mas/pixel) and the double-correlated read-out mode were used.

For the raw data reduction we subtracted a mean dark from all science frames and the flatfield frames, then divided by the normalized dark-subtracted flatfield and subtracted the mean background by using ESO \textit{eclipse\,/\,jitter}.

\begin{figure}[b]
\begin{center}
 \includegraphics[width=5.26in]{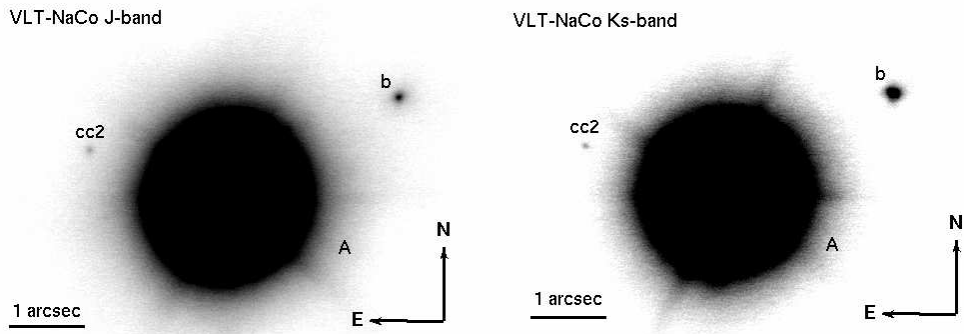}
 \caption{VLT-NaCo J-band and Ks-band images of CT Cha and its 6.3 mag fainter companion candidate (in Ks-band) CT Cha b 2.670\,$\pm$\,0.036'' northwest from March 1st and 2nd 2007. The object marked as 'cc2' was found to be a background object.}
   \label{fig1}
\end{center}
\end{figure}

In all three images a companion candidate is found 2.67'' northwest of CT Cha (Fig.~\ref{fig1}) corresponding to $\sim$\,440 AU at a distance of 165\,$\pm$\,30 pc estimated from the combination of data by \cite[Bertout et al. (1999)]{Bertout99} and \cite[Whittet et al. (1997)]{
Whittet97} for Cha I members.

\subsection{Astrometry}

\begin{figure}[b]
\begin{center}
 \includegraphics[width=2.63in]{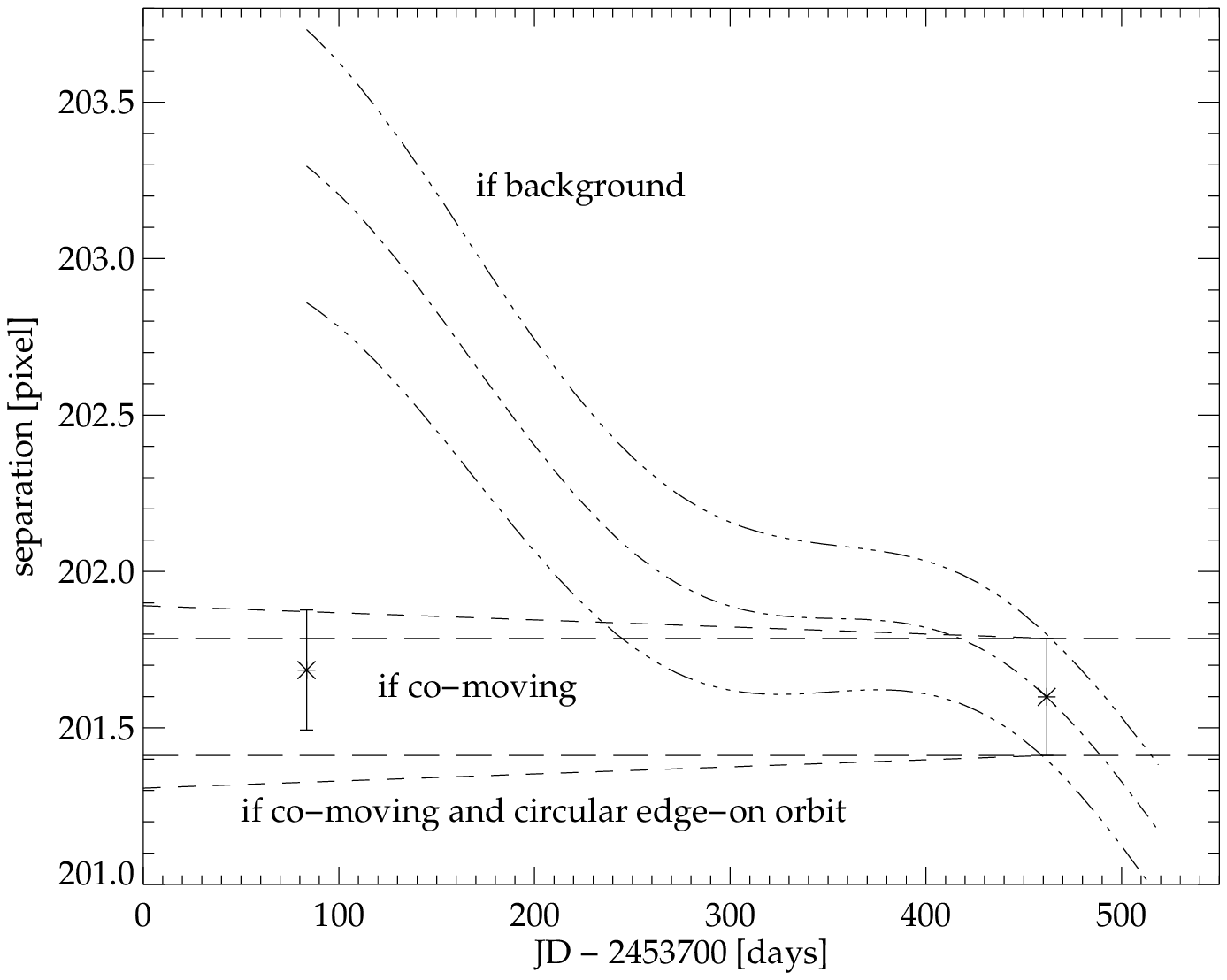}
 \includegraphics[width=2.63in]{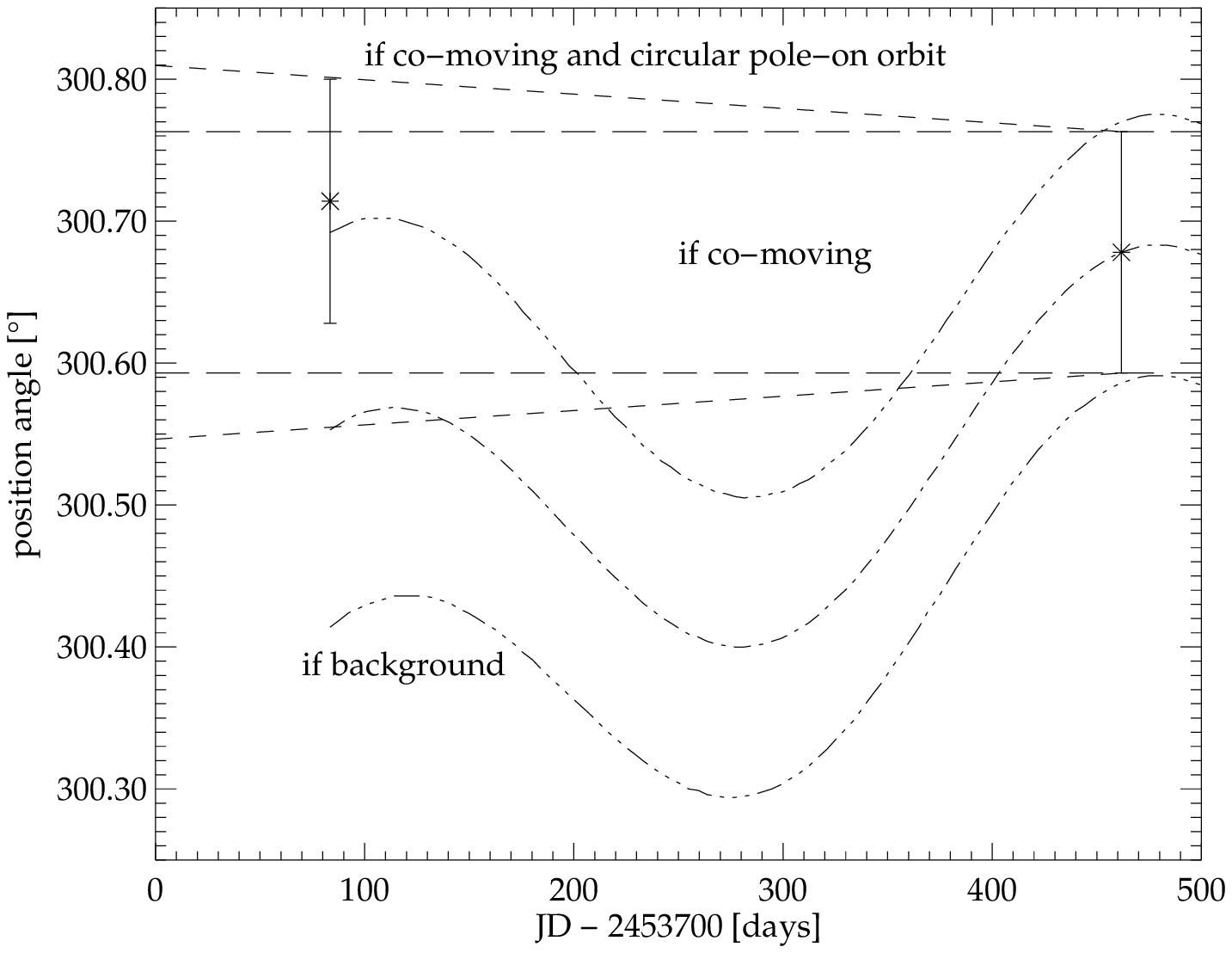}
 \caption{Left: Observed separation between CT Cha A and its companion. Our two measurements from 2006 and 2007 are shown. The long dashed lines enclose the area for constant separation. The dash-dotted line is the change expected if the companion is a non-moving background star. The opening cone enclosed by the dash-double-dotted lines are its estimated errors. The waves of this cone show the differential parallactic motion which has to be taken into account if the other component is a non-moving background star with negligible parallax. The opening short dashed cone is for the combination of co-motion and the maximum possible orbital motion. Right: Same for the position angle (measured from north over east to south) of the companion.}
   \label{fig2}
\end{center}
\end{figure}

To check for common proper motion of the tentative companion of CT Cha we used the proper motion (here PM) of the star published in the literature (table ~\ref{tab1}). We use the weighted mean proper motion for checking, whether the two objects show common proper motion below.

We calibrated the NaCo data using the wide binary star HIP 73357 for our two measurements in 2006 and 2007. The astrometric calibration for our images was done relative, hence the error bars of separation and position angle given in table~\ref{tab2} include possible orbital motion of the binary inbetween our observations as well as their measurement uncertainties.

To determine the positions of both components we constructed a reference PSF from both objects. Thus, we obtained an appropriate reference PSF for each single image. By using IDL/starfinder we scaled and shifted the reference PSF simultaneously to both components in each of our individual images by minimizing the residuals. As a result we obtained positions (see table~\ref{tab2}) and relative photometry (see table~\ref{tab3}) including realistic error estimates for each object by averaging the results of all single images taken within each epoch.

From separation in figure \ref{fig2}, we can exclude by 3.4\,$\sigma$ that CT Cha b is a non-moving background object. Due to the location of the companion (northwest of A) and the proper motion (towards northwest), the position angle information does not give much additional significance (1.0\,$\sigma$ deviation from the background hypothesis), hence resulting in a combined significance of 3.7\,$\sigma$. For a K7 star and a sub-stellar companion (see below) and the given separation (at $\sim165$ pc), the orbital period is $\sim 11000$ yrs, so that the maximum change in separation due to orbital
motion (for circular edge-on orbit) is $\sim 1$ mas/yr or $\sim 0.07$ pix/yr ($\sim 0.03\,^{\circ}$/yr in PA for pole-on orbit).
Neither in separation nor in position angle (measured from north over east to south), being $\sim\,300\,^{\circ}$, signs of orbital motion could be detected given the short epoch difference ($\sim$\,1 year).

\begin{table}
  \begin{center}
  \caption{Proper motions of CT Cha}
  \label{tab1}
 {\scriptsize
  \begin{tabular}{|l|r@{\,$\pm$\,}l|r@{\,$\pm$\,}l|}\hline 
Reference &\multicolumn{2}{c|}{$\mu_{\alpha} \cos{\delta}$}    & \multicolumn{2}{c|}{$\mu_{\delta}$}  \\
          &\multicolumn{2}{c|}{[mas/yr]} & \multicolumn{2}{c|}{[mas/yr]}\\
\hline
UCAC2 \cite[(Zacharias et al. 2004)]{Zacharias04} & -22.2  & 5.2  & 7    & 5.2 \\
ICRF ext. \cite[(Camargo et al. 2003)]{Camargo03}  & -18    & 10   & 4    & 9   \\
\hline
weighted mean                          & -21.3  & 4.6  & 6.3  & 4.5 \\
\hline
  \end{tabular}
  }
 \end{center}
\end{table}

\begin{table}
  \begin{center}
  \caption{Relative astrometric results for CT Cha}
  \label{tab2}
 {\scriptsize
  \begin{tabular}{|l|c|c|r@{\,$\pm$\,}l|}\hline 
Epoch difference & Target & Change in separation  & \multicolumn{2}{c|}{Change in PA$^1$}\\
$[\mathrm{days}]$ & & [pixel] &  \multicolumn{2}{c|}{[$\deg$]} \\
\hline
    378.08299  & CT Cha Ab & -0.086 $\pm$ 0.268 & -0.036 & 0.121\\
\hline
  \end{tabular}
  }
 \end{center}
\vspace{1mm}
 \scriptsize{
 {\it Notes:}\\
All Ks band images.
  $^1$PA is measured from N over E to S. PA is given relative to the first epoch (absolutely measured it is 300.71 $\pm$\,1.24$^{\circ}$) in Feb 2006.\\
}
\end{table}

While the negligible differences seen in PA and separation between the different observations are consistent with common proper motion, a possible difference in proper motion between both objects of up to a few mas/yr cannot be excluded from the data; such a difference in proper motion would be typical for the velocity dispersion in star forming regions like Cha I \cite[(Ducourant et al. 2005)]{Ducourant05}, so that we cannot yet exclude that both objects are independent members of Cha I, but not orbiting each other. Even if this would be the case, age and distance would be the same as assumed below for both objects, hence also the spectral type and its mass.

\subsection{Photometry}

As described in the last section we obtained from the PSF fitting of both components also the relative brightnesses.
Using the photometry of CT Cha A from the Two Micron All Sky Survey (2MASS) catalogue of $J$\,=\,9.715\,$\pm$\,0.024\,mag and $K$\,=\,8.661\,$\pm$\,0.021\,mag and adding 0.3\,mag variability of A estimated from data by \cite[Batalha et al. (1998)]{Batalha98}, \cite[Ghez et al. (1997)]{Ghez97} and \cite[Lawson et al. (1996)]{Lawson96} to the error of b, we obtain the photometry of the companion using our measured brightness difference, see table \ref{tab3}.

The Ks band magnitudes agree within the 1$\sigma$ errors of the two epochs, giving no indication of photometric variability.
From the extinction corrected $J_{0}-Ks_{0}$=\,1.240\,$\pm$\,0.490\,mag (from A$_{V}$ and extinction law by \cite[Rieke \& Lebofsky et al. (1985)]{Rieke85}) we estimate a spectral type of M5 -- L5 for the faint companion CT Cha b using \cite[(Golimowski et al. 2004)]{Golimowski04}.

\begin{table}
  \begin{center}
  \caption{Apparent magnitudes of the companion CT Cha b}
  \label{tab3}
 {\scriptsize
  \begin{tabular}{|l|r@{\,$\pm$\,}l|c|c|}\hline 
Epoch & \multicolumn{2}{c|}{J-band} & Ks-band \\
\hline
17 Feb 2006   & --     & --    & 14.951 $\pm$ 0.302\\
 1/2 Mar 2007 & 16.607 & 0.302 & 14.891 $\pm$ 0.301\\
\hline
  \end{tabular}
  }
 \end{center}
\end{table}

\subsection{Conclusions}

We have found a new co-moving companion candidate of CT Cha and estimated its spectral type from photometry to be M5 -- L5. After this result we have taken integral field spectra of the object with the VLT instrument SINFONI and could determine temperature and surface gravity of the object by comparison with synthetic spectra by \cite[Brott \& Hauschildt (2005)]{Brott05}. From comparison with the eclipsing binary brown dwarf 2M0535 \cite[(Stassun et al. 2006)]{Stassun06} we conclude that the object has less mass than 36 M$_{Jup}$, similar to the GQ Lup companion found by \cite[Neuh\"auser et al. (2005)]{Neuhauser05}. Our best mass estimate from the spectral modelling gives a few Jupiter masses, so that the companion can be a planet imaged directly, hence has to be seen as planetary mass companion candidate, hence called b (small letter for planets and planet candidates). A detailed discussion of the spectra and further informations can be found in \cite[Schmidt et al. (submitted to A\&A)]{Schmidt0708}.

\end{document}